\documentclass[letterpaper, 10 pt, conference]{ieeeconf}  

\IEEEoverridecommandlockouts                              

\usepackage{graphics} 
\usepackage{amsmath} 
\usepackage{amssymb}  
\usepackage{algorithm}
\usepackage{algorithmicx,algpseudocode}
\usepackage{subcaption}
\usepackage{cite}
\usepackage{balance}
\usepackage{color}
\usepackage{hyperref}

\definecolor{tuberlin_red}{cmyk}{0.2,1,1,0}
\definecolor{tuberlin_orange}{cmyk}{0,0.7,0.95,0}
\definecolor{tuberlin_darkgray}{cmyk}{0,0,0,0.8}
\definecolor{tuberlin_lightgray}{cmyk}{0,0,0,0.4}

\newtheorem{assumption}{Assumption}

\title{\LARGE \bf
    Adaptive Meta-Learning-Based KKL Observer Design for Nonlinear Dynamical Systems

\author{Lukas Trommer$^{1}$, Halil Yi\u{g}it Öksüz$^{2}$}

\thanks{$^{1}$ L. Trommer is with the Control Systems Group at the Technical University of Berlin, Germany. {\tt\footnotesize\{lukas.trommer@campus.tu-berlin.de\}}}

\thanks{$^{2}$ H. Y. Öksüz is with the Control Systems Group and Science of Intelligence, Research Cluster of Excellence, at the Technical University of Berlin, Germany. {\tt\footnotesize\{oksuz@tu-berlin.de\}}}%

\thanks{* H. Y. Öksüz acknowledges funding by the Deutsche Forschungsgemeinschaft (DFG, German Research Foundation) under Germany’s Excellence Strategy – EXC 2002/1 “Science of Intelligence” – project number 390523135.}

}

\begin{document}

\maketitle
\thispagestyle{empty}
\pagestyle{empty}

\begin{abstract}
The theory of Kazantzis-Kravaris/Luenberger (KKL) observer design introduces a methodology that uses a nonlinear transformation map and its left inverse to estimate the state of a nonlinear system through the introduction of a linear observer state space. Data-driven approaches using artificial neural networks have demonstrated the ability to accurately approximate these transformation maps. This paper presents a novel approach to observer design for nonlinear dynamical systems through meta-learning, a concept in machine learning that aims to optimize learning models for fast adaptation to a distribution of tasks through an improved focus on the intrinsic properties of the underlying learning problem. We introduce a framework that leverages information from measurements of the system output to design a learning-based KKL observer capable of online adaptation to a variety of system conditions and attributes. To validate the effectiveness of our approach, we present comprehensive experimental results for the estimation of nonlinear system states with varying initial conditions and internal parameters, demonstrating high accuracy, generalization capability, and robustness against noise.
\end{abstract}


\section{INTRODUCTION}

In many control applications, it might not always be feasible to directly measure all states of a dynamical system. In such scenarios, a reliable estimation of these unmeasurable state variables becomes paramount and is usually accomplished through state observers. While the design principles of \textit{Luenberger} observers \cite{luenberger1964observer} are commonly applied to linear systems, addressing this challenge for nonlinear systems is inherently more complex \cite{bernard2022observer}. Numerous efforts have been directed to develop nonlinear observers using extended Kalman filters \cite{zeitz1987extended,reif1999extended,boutayeb1999strong}. However, most of these methods depend on linearization techniques, and therefore, they offer only local convergence.

The theory of \textit{Kazantzis-Kravaris/Luenberger} (KKL) observers \cite{kazantzis1998observer} extends the Luenberger observer design principles to nonlinear systems. KKL observers are founded on the existence of an injective mapping and its left inverse, enabling the transformation of a nonlinear system into a set of first-order linear differential equations subject to the injection of the nonlinear system output. The existing literature provides comprehensive analyses of KKL observers, including the exploration of requirements for the existence of an injective transformation, discussions on the dimensionality of the linear state space \cite{andrieu2006existencekkl,andrieu2014convergence}, and the extension of these principles to non-autonomous systems \cite{bernard2019nonautonomous,ramos2020numerical}. Although it has been demonstrated that the use of an approximation for the state space transformation is sufficient, deriving a concrete formulation of the corresponding mapping and its left inverse for generic problem statements remains a formidable challenge \cite{andrieu2021inversionmaps}. One viable approach is to leverage data-driven methods, involving the collection of data samples through system simulation or measurement. With increased computational power and a growing interest in machine learning, neural networks have emerged as versatile solutions for addressing highly complex and nonlinear problems in a variety of domains \cite{lecun2015deep,goodfellow2016deep}, since they are recognized as universal approximators when appropriately dimensioned \cite{hornik1989universalapproximator}. Various strategies have been proposed to formulate the approximation of KKL observer transformation maps as regression problems solvable through feedforward neural networks. In \cite{ramos2020numerical}, the nonlinear system and the linear part of the observer have been simulated forward in time, starting from pre-specified initial system conditions. Pairwise sets of state data, where one state functions as the input and the other as the label for the nonlinear regression problem, have been generated in order to determine the respective transformation maps. In \cite{peralez2021deep}, a deep auto-encoder has been introduced for systems in discrete time, which learns the transformation maps by enforcing observer dynamics in a latent linear space and introducing a measure of reconstruction loss following the composite application of forward and inverse transformations during training. An unsupervised learning technique has been utilized in \cite{peralez2022neuralkkl}, whose aim is to improve convergence accuracy by using a neural network-based ensemble learning approach. By introducing a physics-informed loss component during training, some improvements in robustness, accuracy, and generalization capabilities have been observed in \cite{niazi2023learningbased}.

It is important to emphasize that learning-based approaches for KKL observer design require labeled, nonlinear system state data for training. While the estimation capabilities of these data-driven approaches benefit from substantial datasets, it is not always practical to utilize arbitrary quantities of data for training. Moreover, in certain scenarios requiring accurate state estimation over varying system parameters, training data may only be available for a limited subset of parameter values. The imposed learning challenge in that case furthermore includes the aim for strong model generalization, allowing accurate estimation even in parameter ranges for which no data is accessible during training.

Aiming to enhance these capabilities, we resort to \textit{Model-Agnostic Meta-Learning} (MAML), which facilitates training a model on a multitude of tasks drawn from a task distribution \cite{finn2017maml,finn2019online,fallah2020convergence} in order to quickly adapt and generalize to new tasks from that distribution with only a few gradient steps, rendering it highly data-efficient for adaptation. We introduce a novel approach to learning-based KKL observer design, harnessing the fundamental principles of MAML to enhance the precision of nonlinear system state estimation through online adaptation to distinct system attributes. The adaptation process exclusively depends on system output measurements to extract inherent system knowledge and dynamically adjust estimation capabilities. This proposed method deviates from the static learning paradigms prevalent in previous methodologies, marking a transition towards enhanced adaptability.

The remainder of the paper is organized as follows. In section \ref{sec:background}, the theory of KKL observers and the fundamental principles of MAML are provided. In Section \ref{sec:methodology}, we present our novel meta-learning-based approach to KKL observer design in detail. We then evaluate the performance of our proposed approach in Section \ref{sec:experiments}, providing empirical results and analysis. Finally, Section \ref{sec:conclusion} offers our concluding remarks.

\subsection*{Notation}

The set of real numbers is denoted by $\mathbb{R}$, $\mathbb{R}^{m}$ represents the $m$-dimensional Euclidean space, $\mathbb{R}_{\geq 0}$ the set of nonnegative real numbers. $\mathbb{N}$ respectively denotes the set of natural numbers. The Euclidean norm of the vector $x\in \mathbb{R}^{m}$ is denoted by $||x||$. The matrix exponential of matrix $A$ is defined as $e^{A} = \sum_{k=0}^{\infty} \frac{1}{k!} A^k$.
Given a finite set $S$, its cardinality is denoted by $|S|$.  The notions $1_N$, $I_N$, and $\mathrm{diag}(1,2,...,N)$ respectively denote the $N$-dimensional vector with all entries equal to one, the $N\times N$ dimensional identity matrix, and the diagonal matrix with entries $1,2,..., N$. $[a,b]^m\subset\mathbb{R}^m$ denotes the $m$-ary Cartesian power of the finite interval $[a,b]\subset\mathbb{R}$.

\section{BACKGROUND}
\label{sec:background}

\subsection{KKL Observers}

We consider a dynamical, autonomous, nonlinear system
\begin{equation}
\label{eq:nonlinear_system}
    \begin{cases}
        \dot{x}(t)&=f\left(x(t)\right)\\
        y(t)&=h\left(x(t)\right)
    \end{cases}
\end{equation}
where $x\in\mathcal{X}\subset\mathbb{R}^{d_x}$ denotes the nonlinear system state in state space $\mathcal{X}$, and $y\in\mathcal{Y}\subset\mathbb{R}^{d_y}$ the system output. $f:\mathcal{X}\mapsto\mathbb{R}^{d_x}$ and
$h:\mathcal{X}\mapsto\mathcal{Y}$ are smooth functions that articulate the nonlinear system dynamics and measurable outputs, respectively. As elucidated in \cite{kazantzis1998observer, andrieu2006existencekkl,andrieu2014convergence
}, the formulation of a KKL observer entails the identification of an injective\footnote{$\forall x_1, x_2\in\mathcal{X}:F(x_1)=F(x_2)\implies x_1=x_2$.} transformation map, denoted by $F:\mathcal{X}\mapsto\mathcal{Z}$, that serves the purpose of converting the dynamics of the nonlinear system (\ref{eq:nonlinear_system}) into a linear dynamical system described by
\begin{equation}
\label{eq:linear_observer}
    \dot{z}(t)=Az(t)+By(t),
\end{equation}
\noindent
where $z(t)\in\mathcal{Z}$ is the linear system state in state space $\mathcal{Z}\subset\mathbb{R}^{d_z}$. Hence, the relationship between the linear state $z(t)$ and the nonlinear state $x(t)$ is expressed as $z(t)=F\left(x(t)\right)$. It is imperative that the transformation map $F$ satisfies the following partial differential equation:
\begin{equation}
\label{eq:linear_observer_pde}
    \frac{\partial}{\partial x}\left(F\left(x(t)\right)\right)f\left(x(t)\right)=AF\left(x(t)\right)+Bh\left(x(t)\right).
\end{equation}

Since the transformation map $F$ is injective, the KKL observer for the nonlinear system (\ref{eq:nonlinear_system}) is instantiated through the application of the inverse transformation map $F^{-1}:\mathcal{Z}\mapsto\mathcal{X}$ to the linear observer state $z(t)$ and is defined as
\begin{equation}
\label{eq:kkl_observer}
    \begin{cases}
        \dot{z}(t)&=Az(t)+By(t)\\
        x(t)&=F^{-1}\left(z(t)\right).
    \end{cases}
\end{equation}
This formulation incorporates the appropriate initialization of the linear observer state with $z(0)=F\left(x(0)\right)$.
In the context of learning-based KKL observers, we approximate $F$ and $F^{-1}$ as depicted by \cite{andrieu2006existencekkl} and denote those as $\hat{F}$ and $\hat{F}^{-1}$. In order to ensure the existence of such transformation maps, we make the following assumptions:
\begin{assumption}
    \label{ass:1}
    Let the state trajectory of (\ref{eq:nonlinear_system}) with the initial state $x(0)=x_0$ be expressed by $x(t;x_0)$.
    There exists a compact set $\mathcal{X}\subset\mathbb{R}^{d_x}$ such that $\forall x_0\in\mathcal{X},t\in\mathbb{R}_{\geq 0}:x(t;x_0)\in\mathcal{X}$.
\end{assumption}
\begin{assumption}
     \label{ass:2}
    There exists an open bounded set $\mathcal{S}\supset\mathcal{X}$ wherein (\ref{eq:nonlinear_system}) is backward $\mathcal{S}$-distinguishable\footnote{Given an open set $\mathcal{S}\supset\mathcal{X}$, (\ref{eq:nonlinear_system}) is said to be \textit{backward $\mathcal{S}$-distinguishable} on $\mathcal{X}$ if for every pair of distinct initial conditions $x_{0,1},x_{0,2}\in\mathcal{X}$, there exists $\tau<0$ such that $x(t;x_{0,1}), x(t;x_{0,2})\in\mathcal{S}$ are well-defined for $t\in[\tau,0]$, and $h(x(\tau;x_{0,1}))\neq h(x(\tau;x_{0,2}))$\cite{niazi2023learningbased}.}.
\end{assumption}

Previous studies such as \cite{andrieu2006existencekkl,bernard2022observer, brivadis2023kklremarks,niazi2023learningbased} emphasize that a uniformly injective\footnote{A function $\rho:\mathbb{R}_{\geq 0}\mapsto\mathbb{R}_{\geq 0}$ is said to be of class $\mathcal{K}$ if it is continuous, zero at zero, and strictly increasing. The map $F$ is said to be \textit{uniformly injective} if there exists a class $\mathcal{K}$ function $\rho$ such that $\forall x_1,x_2\in\mathcal{X}:\|x_1-x_2\|\leq\rho(\|F(x_1)-F(x_2)\|)$.\cite{niazi2023learningbased}} map $F$ satisfying (\ref{eq:linear_observer_pde}) exists if $(A,B)$ is controllable, $A$ is Hurwitz, and  Assumptions \ref{ass:1} and \ref{ass:2} hold.
For formal proof and a more comprehensive context, we direct readers to the cited works.

\subsection{Model-Agnostic Meta-Learning (MAML)}

The concept of MAML \cite{finn2017maml} revolves around training a model on a distribution of tasks, enabling it to rapidly adapt to new tasks drawn from the same distribution. The central notion is to acquire transferable intrinsic features across the given task distribution.

Consider a task $T$ defined as a tuple $T=(u, o)$, where $u\in\mathcal{U}$ denotes the input belonging to the input space $\mathcal{U}$ and $o\in\mathcal{O}$ corresponds to the desired output within the output space $\mathcal{O}$. In the context of meta-learning, we consider a model $\hat{g}_{\theta}\in\mathcal{G}$ parameterized by $\theta$. Here, $\mathcal{G}\subset\{\mathcal{U}\mapsto\mathcal{O}\}$ denotes the function space containing $\hat{g}_{\theta}$, which maps inputs to outputs. Additionally, a loss function $\mathit{L}:\mathcal{O}\times\mathcal{U}\times\mathcal{G}\mapsto\mathbb{R}$ is employed. Tasks are sampled from a given task distribution $\mathcal{T}$, denoted as $T\sim\mathcal{T}$. Throughout this distribution, the tasks vary, exposing the model to different inputs and corresponding desired outputs. During the meta-training process, the model parameters evolve to enable adaptation to specific tasks $T$ using only a limited number of samples and gradient updates.

MAML accomplishes this behavior through a training algorithm characterized by a nested loop. At the beginning of each outer loop iteration, a meta-batch of tasks $\{T_i|T_i\in\mathcal{T}, i=1,2,...,N_\mathrm{batch, meta}\}$ is sampled. Subsequently, an inner loop is entered, where the task-specific model parameters $\theta_i$ are computed by using \textit{adaptation data points} $(u_{i,\mathrm{a}}, o_{i,\mathrm{a}})$ from each task $T_i$ with the update rule
\begin{equation}
    \theta_{i}=\theta-\alpha\nabla_\theta L(o_{i,\mathrm{a}}, u_{i,\mathrm{a}}, \hat{g}_\theta)
\end{equation}
and adaptation learning rate $\alpha$.

Upon completion of the inner loop updates, a meta-update for the model parameters $\theta$ is executed. This update entails the accumulation of losses computed based on a set of predictions for \textit{query data points} $(u_{i,\mathrm{q}}, o_{i,\mathrm{q}})$ which are generated using the updated, task-specific parameter sets $\theta_i$ obtained from the inner loop. Mathematically, this update can be expressed as
\begin{equation}
\label{eq:maml_outer_loop}
    \theta=\theta-\beta\nabla_\theta\sum_{T_i \sim \mathcal{T}} L(o_{i,\mathrm{q}}, u_{i,\mathrm{q}}, \hat{g}_{\theta_i})
\end{equation}
with meta learning rate $\beta$.
The computation of gradients in (\ref{eq:maml_outer_loop}) encompasses second-order gradients through the inner loop gradients. This results in an optimization process where the parameters in the outer loop are adjusted in a manner aimed at minimizing the loss function through consideration of the inner loop parameter updates, which reflect the task-specific adaptation process.

\subsection{Problem Statement}

In this paper, our objective is to find a more general approximation of transformation maps $F$ and $F^{-1}$ given in (\ref{eq:linear_observer_pde}) and (\ref{eq:kkl_observer}) through $\hat{F}_\theta$ and $\hat{F}^{-1}_\eta$, realized as feedforward neural network models with learnable parameters $\theta$ and $\eta$. Given variations in system attributes, such as the initial nonlinear system state $x(0)$ or the intrinsic parameterization of the nonlinear system function $f(\cdot)$, we define a distribution of learning tasks, $\mathcal{T}$, with tasks arising from notable differences in the trajectories $x(t)$ and $z(t)$ of systems (\ref{eq:nonlinear_system}) and (\ref{eq:linear_observer}) over time $t$. In this case, employing a meta-learning approach on the task distribution $\mathcal{T}$ is expected to allow for both, better generalization across $\mathcal{T}$, and to adapt to a single task $T\sim\mathcal{T}$ as it is being observed.

\section{METHODOLOGY}
\label{sec:methodology}

\subsection{Task-Specific Data Generation}

For each task $T\sim\mathcal{T}$, we construct a dataset $\mathcal{D}_T\subset\mathcal{X}\times\mathcal{Z}$ which contains pairs of data samples, denoted as $(x(t;T),z(t;T))$. Here, $x(t;T)$ represents the state trajectory of the nonlinear system related to task $T$, while $z(t;T)$ corresponds to the state of the linear system, respectively. This synthetic approach of data generation, in line with methodologies presented in \cite{ramos2020numerical, niazi2023learningbased}, employs the RK4 method \cite{runge1895diffeq, kutta1901diffeq} (or another suitable solver) to simultaneously forward simulate both systems (\ref{eq:nonlinear_system}) and (\ref{eq:linear_observer}) for a given task $T$. The resulting dataset can be expressed as
\begin{equation}
\label{eq:dataset}
    \mathcal{D}_{T}=\{\left(x(t;T),z(t;T)\right)|t=k\Delta t,k=0,1,...,N\}
\end{equation}
where $\Delta t$ is the time step size used in the solver and $N\in \mathbb{N}$.

Given the analytic solution to the system of linear differential equations in (\ref{eq:linear_observer})\cite{chen1984linear}
\begin{equation}
\label{eq:linear_system_solution}
    z(t)=e^{At}z(0)+\int_0^t e^{A(t-\tau)}By(\tau)\mathrm{d}\tau,
\end{equation}
the effect of any initial state $z(0)\neq F(x(0))$ would eventually diminish due to the Hurwitz property of $A$. However, an arbitrary initialization for training data generation might introduce regression errors, particularly for simulated data samples near $t=0$, depending on the choice of the matrix $A$. To mitigate such errors resulting from inaccurate training data, we adopt a backward sampling method for (\ref{eq:nonlinear_system}) and (\ref{eq:linear_observer}). This strategy, as described in prior works such as \cite{buissonfenet2023gaintuning, niazi2023learningbased}, allows us to obtain the precise initial state $z(0)=F(x(0))$ at which the linear system has reached the steady state:
\begin{enumerate}
    \item Given some small $\epsilon >0$, the system matrix $A$, its eigendecomposition $A=V\Lambda V^{-1}$, and minimum eigenvalue $\lambda_{A,\mathrm{min}}$, compute the time $\tau<0$ such that $||e^{A(-\tau)}z(\tau)||<\epsilon$ using
    \begin{equation}
        \tau\leq\frac{1}{\lambda_{A,\mathrm{min}}}\ln\left(\frac{\epsilon}{\mathrm{cond}(V)\|z(\tau)\|}\right).
    \end{equation}
    \item Simulate $x(t)$ in backward time for $t\in[\tau,0]$ given $x(0)$ and compute $y(t)=h(x(t))$.
    \item Simulate $z(t)$ in forward time for $t\in[\tau,0]$ given $z(\tau)$ and $y(t)$ from the previous step to obtain $z(0)=F\left(x(0)\right)$.
\end{enumerate}

\subsection{Forward Computation During Training}

During the training process, we employ two distinct modes of forward computation. These modes can be categorized as either sequential or parallel in their application of the map approximations $\hat{F}_\theta$ and $\hat{F}^{-1}_\eta$.

The sequential method reconstructs the nonlinear system state approximation, $\hat{x}(t)$, using the linear system state approximation, $\hat{z}(t)$, as the basis for computation. This is achieved through
\begin{equation}
\label{eq:sequential_forward_computation}
    \hat{z}(t)=\hat{F}_\theta\left(x(t)\right),\;\;\;\hat{x}(t)=\hat{F}^{-1}_\eta\left(\hat{F}_\theta\left(x(t)\right)\right).
\end{equation}

In contrast, the parallel method avoids the composition of both map approximations and directly utilizes labels for the state of the linear observer system $z(t)$ to compute $\hat{x}(t)$. The forward computation in this mode can be expressed as
\begin{equation}
\label{eq:parallel_forward_computation}
    \hat{z}(t)=\hat{F}_\theta\left(x(t)\right),\;\;\;\hat{x}(t)=\hat{F}^{-1}_\eta\left(z(t)\right).
\end{equation}

\begin{figure}
    \centering
    \includegraphics{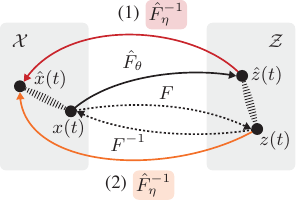}
    \caption{Modes of forward computation during training, sequential (1) and parallel (2).}
    \label{fig:forward_computation}
\end{figure}

The sequential approach, depicted in Figure \ref{fig:forward_computation} (1), applies $\hat{F}^{-1}_\eta$ to the linear state estimate $\hat{z}(t)$, aiming to provide a direct inverse to the approximation $\hat{F}_\theta$. Conversely, the parallel approach, illustrated in Figure \ref{fig:forward_computation} (2), applies $\hat{F}^{-1}_\eta$ independently from $\hat{F}_\theta$ to directly invert $F$.

\subsection{Mixed-Task Learning Methods}

We consider a set of training tasks $\mathcal{T}_\mathrm{train}$ sampled from the distribution $\mathcal{T}$ with task-specific datasets $\mathcal{D}_T$ for $T\in\mathcal{T}_\mathrm{train}$ and a joint dataset $\mathcal{D}=\cup_{T\in\mathcal{T}_\mathrm{train}}\mathcal{D}_T$, which comprises mixed data points corresponding to different tasks in $\mathcal{T}_\mathrm{train}$. We address the supervised regression problem to learn $\hat{F}_\theta$ and $\hat{F}^{-1}_\eta$ using (\ref{eq:parallel_forward_computation}) on mixed data points from $\mathcal{D}$ \cite{ramos2020numerical}. This approach, termed \textit{parallel mixed-task learning}, involves the loss functions
\begin{align}
    \begin{split}
        L_z\left(z(t),x(t),\hat{F}_\theta\right)&=\left\|z(t)-\hat{F}_\theta\left(x(t)\right)\right\|^2\\
        L_x\left(x(t),z(t),\hat{F}^{-1}_\eta\right)&=\left\|x(t)-\hat{F}^{-1}_\eta\left(z(t)\right)\right\|^2,
    \end{split}
\end{align}
which are independently addressed with the optimization problems
\begin{equation}
    \min_\theta L_z\left(z(t),x(t),\hat{F}_\theta\right)
\end{equation}
and
\begin{equation}
    \min_\eta L_x\left(x(t),z(t),\hat{F}^{-1}_\eta\right).
\end{equation}

Similarly, following the methodology proposed in \cite{niazi2023learningbased} based on the references therein, we denote the same supervised regression problem when using the sequential forward computation of (\ref{eq:sequential_forward_computation}). We refer to this approach as \textit{sequential mixed-task learning}. Induced by the composition of both maps, the loss function $L_x$ in this case becomes
\begin{equation}
    L_x\left(x(t),x(t),\hat{F}^{-1}_\eta\circ\hat{F}_\theta\right)=\left\|x(t)-\hat{F}^{-1}_\eta\left(\hat{F}_\theta\left(x(t)\right)\right)\right\|^2,
\end{equation}
which leads to the optimization problem for $\theta$ depending on both $L_x$ and $L_z$ with
\begin{equation}\min_\theta\left(L_x\left(x(t),x(t),\hat{F}^{-1}_\eta\circ\hat{F}_\theta\right)+L_z\left(z(t),x(t),\hat{F}_\theta\right)\right).
\end{equation}
The optimization problem for $\eta$ is then given by
\begin{equation}
    \min_\eta L_x\left(x(t),x(t),\hat{F}^{-1}_\eta\circ\hat{F}_\theta\right).
\end{equation}

\subsection{Meta-Learning for System Output Adaptation}

\begin{algorithm}[t]
    \caption{Meta-Learning for System Output Adaptation}
    \label{alg:meta_learning_system_output_adaption}
    \begin{algorithmic}[1]
        \Require $\mathcal{T}$ (Distribution of learning tasks)
        \Require $\beta$ (Meta learning rate)
        \State Initialize model $\hat{F}^{-1}_\eta$ with parameters $\eta$
        \State Initialize adaptation learning rate $\alpha$
        \While{not done}
            \For{$i\gets 1,N_\mathrm{batch,meta}$}
                \State $\eta_i\gets\eta$ (Initialize task-specific parameters)
                \State $T_i\sim\mathcal{T}$ (Sample learning task)
                \State Generate $\mathcal{D}_{T_i}$
                \For{$j\gets 1,N_\mathrm{adapt}$}
                    \State $(x_{i,\mathrm{a}},z_{i,\mathrm{a}})\sim\mathcal{D}_{T_i}$ (Sample adaptation data)
                    \State $\eta_i\gets\eta_i-\alpha\nabla_{\eta_i} L_y(h(x_{i,\mathrm{a}}), z_{i,\mathrm{a}}, h\circ \hat{F}^{-1}_{\eta_i})$
                \EndFor
                \State $(x_{i,\mathrm{q}},z_{i,\mathrm{q}})\sim\mathcal{D}_{T_i}$ (Sample query data)
            \EndFor
            \State $\eta\gets\eta-\beta\nabla_\eta\sum_{i} L_x(x_{i,\mathrm{q}}, z_{i,\mathrm{q}}, \hat{F}^{-1}_{\eta_i})$
            \State $\alpha\gets\alpha-\beta\nabla_\alpha\sum_{i} L_x(x_{i,\mathrm{q}}, z_{i,\mathrm{q}}, \hat{F}^{-1}_{\eta_i})$
        \EndWhile
        \State \Return $\hat{F}^{-1}_\eta$
    \end{algorithmic}
\end{algorithm}

\begin{figure}
    \centering
    \includegraphics{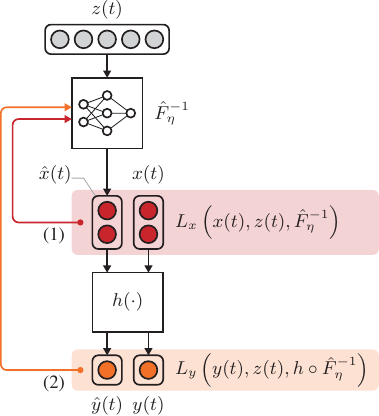}
    \caption{Computational structure of meta-learning for system output adaptation algorithm. (1) The approximation loss $L_x$ for the entire system state vector is computed for meta-updates only. (2) The approximation loss $L_y$ for the measurable system output contributes to the adaptation updates.}
    \label{fig:meta_learning_structure}
\end{figure}

We introduce the concept of \textit{meta-learning for system output adaptation} as a complementary approach to parallel mixed-task learning if $h(\cdot)$ in (\ref{eq:nonlinear_system}) is known or can be approximated through another data-driven regression approach. We employ the mixed learning method for the approximation $\hat{F}_\theta$ and propose Algorithm \ref{alg:meta_learning_system_output_adaption} to learn the inverse map approximation $\hat{F}^{-1}_\eta$. Inspired by the principles of MAML \cite{finn2017maml}, our meta-learning algorithm follows a nested training loop structure, where inner loop iterations involve multiple \textit{adaptation updates}, while outer loop iterations include \textit{meta-updates}. During each outer iteration, a meta batch of $N_\mathrm{batch,meta}$ learning tasks $T_i$ is sampled from $\mathcal{T}$. It is important to emphasize that new tasks $T_i$ are sampled from $\mathcal{T}$ during each training iteration. However, to maintain consistency with the previously introduced mixed-task learning algorithms, we continue to refer to this set of tasks over the entire execution of the training algorithm collectively as $\mathcal{T}_\mathrm{train}$. Consequently, we ensure that all learning algorithms are applied to the same training data $\mathcal{D}$ to preserve comparability. For each task $T_i$, the inner loop performs $N_\mathrm{adapt}$ optimization steps on a local set of model parameters $\eta_i$ for adaptation data points $(x_{i,\mathrm{a}},z_{i,\mathrm{a}})\sim\mathcal{D}_{T_i}$, utilizing $h(\cdot)$ on estimations $\hat{x}(t)$, the system output loss function defined as
\begin{multline}
\label{eq:update_loss_function}
    L_y\left(y(t), z(t), h\circ\hat{F}^{-1}_\eta\right)=\left\|y(t)-h\left(\hat{F}^{-1}_\eta\left(z(t)\right)\right)\right\|^2
\end{multline}
and adaptation learning rate $\alpha$.
Subsequently, the task-specific model parameters $\eta_i$ are evaluated using query data points $(x_{i,\mathrm{q}},z_{i,\mathrm{q}})\sim\mathcal{D}_{T_i}$, and the approximation loss for predicting the nonlinear system state is accumulated across all tasks $T_i$ within the meta batch. In the next step, the accumulated loss is used to perform an optimization step in the outer loop, updating the model parameters $\eta$ based on the meta-learning rate $\beta$.
In addition to meta-learning the parameters $\eta$, the adaptation learning rate $\alpha$ is also introduced as a trainable parameter during the meta-optimization process. This inclusion enhances accuracy for the specified learning problem and the chosen number of adaptation steps, eliminating the need for manual hyperparameter tuning.
In essence, the described algorithm seeks to determine a set of model parameters $\eta$ that enable the model to be updated during the adaptation procedure of the inner loop. The second-order gradients involved in the meta update process adjust $\eta$ such that task-specific adaptation steps minimize the overall approximation loss $L_x$. Importantly, as the adaptation updates are computed based on the measurable system output, this approach supports online adaptation if a system output estimate $\hat{y}(t)$ can be derived from the estimated system state $\hat{x}(t)$ following the previously mentioned constraint that $h(\cdot)$ or an approximation is known. The computational structure of meta-learning for system output adaptation is further illustrated in Figure \ref{fig:meta_learning_structure}.

\subsection{Online Adaptation}
\label{sec:methodology_online_adaptation}

\begin{figure}
    \centering
    \includegraphics{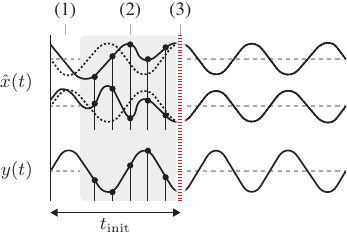}
    \caption{Online adaptation with illustrated sampling delay (1), sampling period (2) and adaptation updates for inverse transformation model parameters (3).}
    \label{fig:adaptation_principle}
\end{figure}

To adapt the meta-learned inverse map approximation $\hat{F}^{-1}_\eta$ online to the specific attributes of the observed nonlinear system, we implement the inner loop adaptation updates from Algorithm \ref{alg:meta_learning_system_output_adaption} using data samples drawn from both $y(t)$ and the estimation $\hat{x}(t)$ governed by the unadapted meta-learning-based KKL observer. It is imperative to emphasize the necessity of an initialization period, denoted as $t_\mathrm{init}$, which arises from this sampling operation, as depicted in Figure \ref{fig:adaptation_principle}.

The minimum duration of $t_\mathrm{init}$ is determined by the number of data points processed during task-specific adaptation and is expressed as:
\begin{equation}
t_\mathrm{init}\geq N_\mathrm{batch}\cdot N_\mathrm{adapt}\cdot\Delta t,
\end{equation}
where $N_\mathrm{batch}$ represents the data batch size. The sampling period can be modified by either delaying, extending (which includes randomly sampling the required data points from within), or using a combination of both strategies. Consequently, the impact of the chosen sampling period on the estimation accuracy additionally relies on the linear system dynamics of the KKL observer and the initial state of (\ref{eq:linear_observer}) determined through $\hat{F}_\theta$.

\section{EXPERIMENTS}
\label{sec:experiments}

Building upon experimental setups given in the literature\cite{ramos2020numerical,peralez2021deep,peralez2022neuralkkl,buissonfenet2023gaintuning,niazi2023learningbased}, we assess the accuracy and performance of our proposed meta-learning algorithm for KKL observer by design employing a variation of the Duffing oscillator as nonlinear system, which is defined as follows:
\begin{equation}
    \begin{cases}
        \dot{x}(t)=
        \begin{bmatrix}
            \dot{x}_1(t)\\\dot{x}_2(t)
        \end{bmatrix}=\lambda
        \begin{bmatrix}
            x_2^3(t)\\-x_1(t)
        \end{bmatrix}+w_x(t)\\
        y(t)=x_1(t)+w_y(t).
    \end{cases}
\end{equation}
The variable $\lambda$ denotes an internal parameter that influences the oscillation frequency and the terms $w_x\in\mathbb{R}^{d_x}$ and $w_y\in\mathbb{R}^{d_y}$ account for system and measurement noise, respectively.

The transformation maps $\hat{F}_\theta$ and $\hat{F}^{-1}_\eta$ are approximated through feedforward neural networks with 5 hidden layers of 50 neurons each, ReLU activation functions and normalization layers. This configuration closely follows the framework presented by \cite{niazi2023learningbased}. We design the linear system (\ref{eq:linear_observer}) of the KKL observer by adopting $A=-\mathrm{diag}(1,2,...,d_z)$, $B=1_{d_z}$ and $d_z=2d_x+1=5$. It is worth noting that, as pointed out by \cite{niazi2023learningbased}, various adjustments such as an $\mathcal{H}_\infty$ design of matrices $A$ and $B$, the generation of an arbitrary amount of training data through simulation, or the use of more complex neural network architectures, could potentially enhance accuracy, generalization capability, and robustness against noise. However, we aim to demonstrate the performance of the previously introduced methodologies under conditions where these improvements may be impractical or unattainable. In the following, we evaluate 4 different learning methods\footnote{The code implementation for the experiments described in this paper is available at \href{https://github.com/lukastrm/metakkl}{\texttt{github.com/lukastrm/metakkl}}.}: sequential mixed-task learning, physics-informed sequential mixed-task learning as introduced by \cite{niazi2023learningbased} as \textit{supervised PINN}, parallel mixed-task learning, and our proposed method, meta-learning for system output adaptation. We use Adam optimization \cite{kingma2015adam} for the former 3 methods as well as the meta-updates of the meta-learning approach and stochastic gradient-descent optimization for the adaptation updates. The proposed meta-learning approach not only functions as an independent learning algorithm but also exhibits improved performance when the inverse map is pre-trained in a parallel mixed-task learning setting. This strategy is incorporated into all our experiments.

To quantify the accuracy of our estimations for $x(t)$ and $z(t)$, we use normalized estimation errors as follows:
\begin{equation}
    e^*_x(t)=\frac{\|x(t)-\hat{x}(t)\|}{\|x(t)\|},\;\;\;
    e^*_z(t)=\frac{\|z(t)-\hat{z}(t)\|}{\|z(t)\|}.
\end{equation}
We furthermore define the normalized estimation error averaged over a set of tasks $\mathcal{T}$ as
\begin{equation}
\label{eq:norm_estimation_error_task_mean}
    \Bar{e}^*_\mathcal{T}(t,e^*,\mathcal{T})=\frac{1}{|\mathcal{T}|}\sum_{T\in\mathcal{T}}e^*(t;T),
\end{equation}
and the normalized estimation error for a task $T\in\mathcal{T}$ averaged over time as
\begin{equation}
\label{eq:norm_estimation_error_time_mean}
    \Bar{e}_t^*(T,e^*)=\frac{1}{N}\sum_{k=0}^N e^*(k\Delta t;T),
\end{equation}
where $e^*(t)$ refers to either $e^*_x(t)$ or $e^*_z(t)$, depending on which state space is under evaluation.

We determine these error metrics for a KKL observer with a linear system that has reached the steady state. For online adaptation, we choose the minimum sampling period which also includes drawing the adaptation samples for the meta-trained KKL observer from the non-steady phase of $z(t)$, but also consider other sampling periods in section \ref{sec:experimentation_x_init_variation}.

\subsection{System Parameter Variation}
\label{sec:experimentation_lambda_variation}

In this experiment, we assess the accuracy of the KKL observer across a distribution of tasks $\mathcal{T}_\lambda$ that incorporate varying values of the parameter $\lambda$. The initial nonlinear system state is fixed at $x(0)=[0.5, 0.5]^T$. During training, we employ a small set of 5 distinct parameter values within the range $\lambda\in[1,5]$ to generate state trajectories. For validation, we extend our analysis to a set of tasks $\mathcal{T}_{\lambda,\mathrm{val}}$, comprising 200 different parameter values for $\lambda\in[1,5]$ within the training range.

\begin{figure}[t]
    \setlength\belowcaptionskip{0.5\baselineskip}
    \centering
    \begin{subfigure}[t]{0.48\columnwidth}
        \centering
        \includegraphics{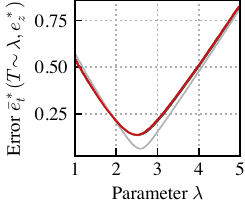}
    \end{subfigure}%
    \hfill
    \begin{subfigure}[t]{0.48\columnwidth}
        \centering
        \includegraphics{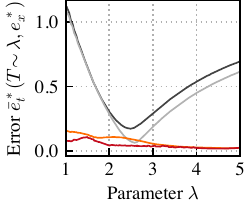}
    \end{subfigure}
    \par
    \begin{subfigure}{\columnwidth}
        \centering
        \includegraphics{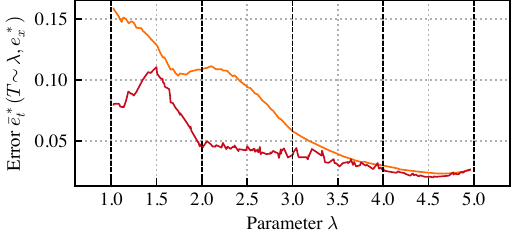}
    \end{subfigure}
    \caption{Simulation results for task distribution $\mathcal{T}_\lambda$ over varying $\lambda$ for different learning methods: (\textcolor{tuberlin_red}{$\bullet$}) meta-learning with online adaptation, (\textcolor{tuberlin_orange}{$\bullet$}) parallel mixed-task learning, (\textcolor{tuberlin_darkgray}{$\bullet$}) sequential mixed-task learning, (\textcolor{tuberlin_lightgray}{$\bullet$}) physics-informed, sequential mixed-task learning. The dashed, black vertical lines in the lower plot indicate the 5 distinct parameter values for $\lambda$ on which the KKL observer was trained.}
    \label{fig:param_error_lmbda}
\end{figure}

The results, depicted in Figure \ref{fig:param_error_lmbda}, reveal two significant findings. Firstly, it is evident that the approximation $\hat{F}_\theta$ displays notable inaccuracies when trained on a dataset involving mixed tasks with varying $\lambda$. This, in turn, results in substantial estimation errors for the KKL observer with sequential mixed-task learning, primarily due to the inverse map approximation $\hat{F}^{-1}_\eta$ being trained on an imprecise representation of the linear system space. In contrast, parallel mixed-task learning and in particular also our meta-learning approach, utilizing parallel forward computation during training, overcome this limitation. The inverse map $\hat{F}^{-1}_\eta$ is consistently trained on accurately simulated data samples for $z(t)$, thus achieving independence from the accuracy of $\hat{F}_\theta$. Secondly, we emphasize how, despite having a limited amount of training data encompassing state trajectories for only 5 distinct values of $\lambda$, our proposed approach demonstrates an improvement in overall validation accuracy through enhanced generalization. Despite the inherent challenge of achieving high accuracy across the entire parameter range due to limited training data, meta-learning and online adaptation leverage the intrinsic system properties acquired from the system output. This allows for more precise estimation of the system state, even for previously unexplored parameter ranges.

\begin{figure*}[t]
    \setlength\belowcaptionskip{0.5\baselineskip}
    \centering
    \begin{subfigure}[t]{0.32\textwidth}
        \centering
        \includegraphics{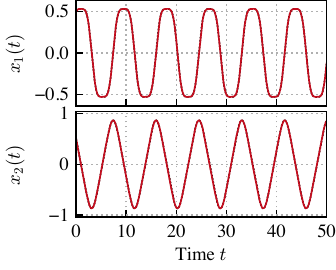}
        \caption{$\hat{x}(t)$ (\textcolor{tuberlin_red}{$\bullet$}, solid) and $x(t)$ (\textcolor{black}{$\bullet$, dotted}) with $x(0)=[0.5,0.5]^T,w_x(t)=0$ and $w_y(t)=0$.}
        \label{fig:state_x_init}
    \end{subfigure}%
    \hfill
    \begin{subfigure}[t]{0.32\textwidth}
        \centering
        \includegraphics{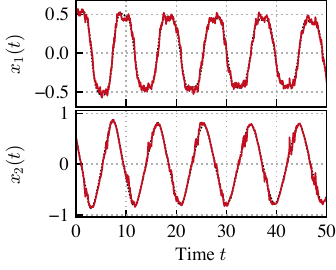}
        \caption{$\hat{x}(t)$ (\textcolor{tuberlin_red}{$\bullet$}, solid) and $x(t)$ (\textcolor{black}{$\bullet$, dotted}) with $x(0)=[0.5,0.5]^T,w_x(t)\sim\mathcal{N}(0,0.1)$ and $w_y(t)\sim\mathcal{N}(0,0.1)$.}
        \label{fig:state_x_init_noise}
    \end{subfigure}%
    \hfill
    \begin{subfigure}[t]{0.32\textwidth}
        \centering
        \includegraphics{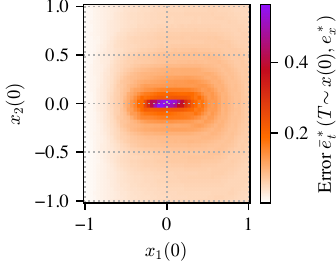}
        \caption{Error profile over $x(0)\in[-1,1]^2$.}
        \label{fig:error_map_x_init}
    \end{subfigure}
    \par
    \begin{subfigure}[t]{0.49\textwidth}
        \centering
        \includegraphics{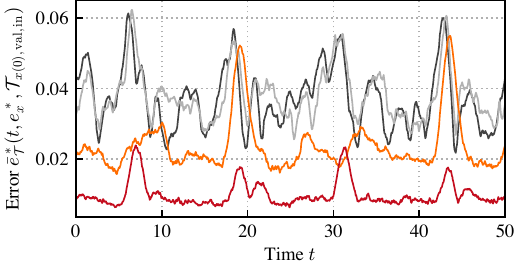}
        \caption{Error for validation tasks within training range ($\mathcal{T}_{x(0),\mathrm{val,in}}$).}
        \label{fig:error_x_init_in}
    \end{subfigure}%
    \hfill
    \begin{subfigure}[t]{0.49\textwidth}
        \centering
        \includegraphics{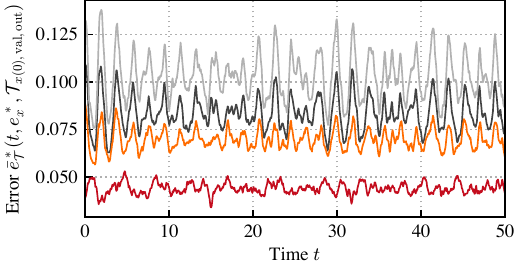}
        \caption{Error validation tasks outside of training range ($\mathcal{T}_{x(0),\mathrm{val,out}}$).}
        \label{fig:error_x_init_out}
    \end{subfigure}
    \caption{Simulation results for task distribution $\mathcal{T}_{x(0)}$ over varying $x(0)$ for different learning methods: (\textcolor{tuberlin_red}{$\bullet$}) meta-learning with online adaptation, (\textcolor{tuberlin_orange}{$\bullet$}) parallel mixed-task learning, (\textcolor{tuberlin_darkgray}{$\bullet$}) sequential mixed-task learning, (\textcolor{tuberlin_lightgray}{$\bullet$}) physics-informed, sequential mixed-task learning.}
\end{figure*}

\subsection{Initial Nonlinear System State Variation}
\label{sec:experimentation_x_init_variation}

In this set of experiments, we assess the accuracy of the KKL observer for a distribution of tasks $\mathcal{T}_{x(0)}$ with varying initial nonlinear system states $x(0)$ and fixed parameter value $\lambda=1$. For training, we draw a set of 50 different initial states $x(0)\in\mathcal{X}_{x(0),\mathrm{train}}\sim[-1,1]^2$ using latin hypercube sampling \cite{iman1999lhs} to generate state trajectories. For validation, we consider two different sets of tasks:
\begin{itemize}
    \item $\mathcal{T}_{x(0),\mathrm{val,in}}$ for 50 different initial states within the training range, i.e. $x(0)\in[-1,1]^2\setminus\mathcal{X}_{x(0),\mathrm{train}}$
    \item $\mathcal{T}_{x(0),\mathrm{val,out}}$ for 80 different initial states outside of the training range with $x(0)\in[-2,2]^2\setminus[-1,1]^2$
\end{itemize}
To evaluate accuracy under the influence of noise, we introduce Gaussian noise components with $w_x(t)\sim\mathcal{N}(0, 0.1)$ and $w_y(t)\sim\mathcal{N}(0, 0.1)$.

Figures \ref{fig:state_x_init} and \ref{fig:state_x_init_noise} display the estimated nonlinear system state $\hat{x}(t)$ using the meta-learning-based KKL observer with online adaptation, clearly highlighting its ability to deliver accurate state estimations, even in the presence of noise.

In Figure \ref{fig:error_x_init_in}, we present the error (\ref{eq:norm_estimation_error_task_mean}) for the validation task set associated with initial system states within the training range. Notably, our proposed methodology consistently achieves the highest accuracy estimations on average when compared to other learning methods. Furthermore, our approach demonstrates strong generalization capabilities, evidenced by the notably low estimation error for initial system states outside of the training range, as portrayed in Figure \ref{fig:error_x_init_out}.

Figure \ref{fig:error_map_x_init} shows the profile of the error (\ref{eq:norm_estimation_error_time_mean}) across the entire training range. In can be observed that the normalized error is larger for initial system states around $x(0)=[0,0]^T$, which can be contributed to numerical imprecisions that arise from a wide normalization range in combination with an absolute loss function during training. This effect has previously also been observed in \cite{ramos2020numerical}.

\begin{figure}
    \centering
    \begin{subfigure}{\columnwidth}
        \centering
        \includegraphics{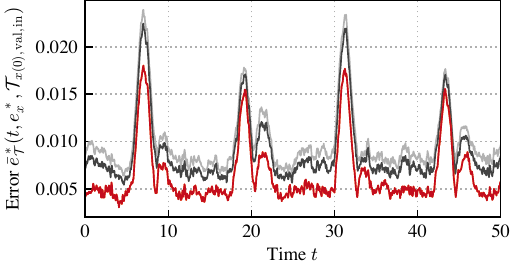}
    \end{subfigure}
    \caption{Simulation results for task distribution $\mathcal{T}_{x(0)}$ over varying $x(0)$ for meta-learned KKL observer with different sampling periods for online adaptation: (\textcolor{tuberlin_lightgray}{$\bullet$}) minimum sampling period without delay, (\textcolor{tuberlin_darkgray}{$\bullet$}) sampling period of size $t=50$ and random sampling of data points within, (\textcolor{tuberlin_orange}{$\bullet$}) minimum sampling period delayed by $-\tau$, (\textcolor{tuberlin_red}{$\bullet$}) sampling period of size $t=50$ delayed by $-\tau$.}
    \label{fig:error_x_init_sampling}
\end{figure}

We further explore the impact of the chosen adaptation sampling period on the precision of the KKL observer. Figure \ref{fig:error_x_init_sampling} presents the simulation results for the meta-learning-based KKL observer with online adaptation, employing four different variations of the sampling strategy. The results highlight that, in the case of the periodic Duffing oscillator, opting for the minimum sampling window without delay leads to the least accurate estimation outcomes. As the linear system error, originating from the initialization using $\hat{F}_\theta$, diminishes over time, expanding the sampling window contributes to improved accuracy. This improvement is due to estimations for $\hat{x}(t)$ being derived from progressively less erroneous samples in $\mathcal{Z}$. Notably, the most accurate estimation performance is observed for delayed sampling windows, ensuring that any error stemming from an inaccurate initial state in $\mathcal{Z}$ has dissipated before the initial adaptation samples are drawn.

\section{CONCLUSION}
\label{sec:conclusion}

In this paper, we have introduced a novel meta-learning-based algorithm to approximate the inverse transformation map, making it adaptable to different system attributes, and conducted a comprehensive comparison of various learning-based KKL observer approaches. We have demonstrated through experiments that our proposed method, integrated with the parallel mixed-task learning algorithm, exhibits heightened estimation accuracy for scenarios both within and outside the training range. This underscores the enhanced generalization capability of meta-learning. Our methodology has also displayed adaptability over varying system parameters, even when only limited task-specific data is available. The integration of meta-learning principles, leveraging the ability to comprehend intrinsic features of the addressed task distribution, has exhibited effectiveness, notably in situations where learning the transformation map $F$ proves challenging within a given setup. This especially outpaces approaches incorporating sequential forward computation during training. 

Our future research will focus on integrating various enhancements, such as those outlined by the authors of \cite{antoniou2018howtomaml}, which could potentially enhance accuracy and foster a more stable learning behavior. Additionally, the necessity of an adaptation sampling period requires careful evaluation. Meta-learning-based KKL observers for systems with dynamically changing attributes may benefit from periodic re-adaptation, but the frequency of attribute changes is constrained by the initialization time of the observer.

\vspace{0.5cm}


\balance
\bibliographystyle{IEEEtran}
\bibliography{ref.bib}

\end{document}